\documentclass[pra,twocolumn,10pt,aps]{revtex4-1}
\usepackage[utf8]{inputenc}  %Input what you want e.g., é, ł, a, ü
\usepackage[T1]{fontenc}     %Output what you want e.g., é, ł, a, ü
\usepackage[british]{babel}  %Do hyphenation according to british english
\usepackage[scaled=0.86]{berasans}  % URL font that go well wtih palatino
\usepackage[colorlinks=true,allcolors=blue]{hyperref}  %Hyperlinks (pink, green, blue)
\usepackage{graphicx} % Package to insert external figures
\usepackage[babel]{microtype}  %Improves text justification
\usepackage{amsmath,amssymb,amsthm,bm,amsfonts,mathrsfs,bbm} %Useful math packages
\usepackage{setspace}
\usepackage{braket}
\usepackage[bottom]{footmisc}
\usepackage{blochsphere}
\usepackage{pgfplots}
\usetikzlibrary{calc,3d,shapes, pgfplots.external, intersections}

\usepackage[bitstream-charter]{mathdesign}
\usepackage{tikz}
\usepackage{mathtools}
\usetikzlibrary{arrows.meta}
\usepackage{csquotes}

\usepackage{xspace}  %Useful to add space in macros
\usepackage{pgfplots}
\usepackage{verbatim}

\newcommand\id{\leavevmode\hbox{\small1\kern-3.3pt\normalsize1}}

\begin{document}

%\preprint{APS/123-QED}

\title{Broadcasting of quantum correlations in qubit-qudit systems}
\author{Rounak Mundra$^{1}$, Dhrumil Patel$^{1}$, Indranil Chakrabarty$^{2}$, Nirman Ganguly$^{3}$ and Sourav Chatterjee$^{1,4}$}
%\email{souravchat25@gmail.com}
\affiliation{$^{1}$Center for Computational Natural Sciences and Bioinformatics, International Institute of Information Technology-Hyderabad, Gachibowli, Telangana-500032, India.\\
$^{2}$Center for Security, Theory and Algorithmic Research, International Institute of Information Technology-Hyderabad, Gachibowli, Telangana-500032, India.\\
$^{3}$ Department of Mathematics, Birla Institute of Technology and Science Pilani, Hyderabad Campus, Telangana-500078, India. \\
$^{4}$SAOT, Erlangen Graduate School in Advanced Optical Technologies, Paul-Gordan-Strasse 6, 91052 Erlangen, Germany.
}

% \author{Dhrumil Patel}
% \email{dhrumil.patel@research.iiit.ac.in}
% \affiliation{Center for Computational Natural Sciences and Bioinformatics, International Institute of Information.
% Technology-Hyderabad, Gachibowli, Telangana-500032, India.}

% \author{Indranil Chakrabarty}
% \email{indranil.chakrabarty@iiit.ac.in}
% \affiliation{Center for Security, Theory and Algorithmic Research, International Institute of Information Technology-Hyderabad, Gachibowli, Telangana-500032, India.\\
% Centre for Theoretical Physics, Jamia Milia Islamia University,Jamia Nagar, Okhla, New Delhi, Delhi 110025.}

% \author{Sourav Chatterjee}
% \email{souravchat25@gmail.com}
% \affiliation{Center for Computational Natural Sciences and Bioinformatics, International Institute of Information Technology-Hyderabad, Gachibowli, Telangana-500032, India.\\}

\begin{abstract}
Quantum mechanical properties like entanglement, discord and coherence act as fundamental resources in various quantum information processing tasks. Consequently, generating more resources from a few, typically termed as broadcasting is a task of utmost significance. One such strategy of broadcasting is through the application of cloning machines. In this article, broadcasting of quantum resources beyond $2 \otimes 2$ systems is investigated. In particular, in $2\otimes3$ dimension, a class of states not useful for broadcasting of entanglement is characterized for a choice of optimal universal Heisenberg cloning machine. The broadcasting ranges for maximally entangled mixed states (MEMS) and two parameter class of states (TPCS) are obtained to exemplify our protocol. A significant derivative of the protocol is the generation of entangled states with positive partial transpose in $3 \otimes 3$ dimension and states which are absolutely separable in $2 \otimes 2$ dimension. Moving beyond entanglement, in $2 \otimes d$ dimension, the impossibility to optimally broadcast quantum correlations beyond entanglement (QCsbE) (discord) and quantum coherence ($l_{1}$-norm) is established. However, some significant illustrations are provided to highlight that non-optimal broadcasting of QCsbE and coherence are still possible. \\

\begin{comment}
Quantum entanglement, quantum correlations and quantum coherence are not only key ingredients but also act as resources for a wide range of quantum computing and information processing tasks. Consequently, generating more resources from a few is a task of utmost significance and is known as the broadcasting of quantum resources. One such strategy of broadcasting is through the application of cloning machines. In this article, we investigate broadcasting of quantum resources beyond $2 \otimes 2$ systems. In particular, in $2\otimes3$ dimensions, we are able characterize a class of states not useful for broadcasting of entanglement for a choice of optimal universal Heisenberg cloning machine. Additionally, we find the broadcasting ranges for maximally entangled mixed states (MEMS) and two parameter class of states (TPCS) as an example. As a significant derivative of this process, we also generate entangled states with positive partial transpose and states which are absolutely separable. Other than entanglement, in $2\otimes d$ dimension, we show that it is impossible to optimally broadcast quantum correlations beyond entanglement (QCsbE) (discord) and quantum coherence ($l_{1}$-norm). Nevertheless, through same examples from $2\otimes3$ systems, we show that non-optimal broadcasting of QCsbE and coherence are still possible.
\end{comment}

\end{abstract}

\maketitle

%\tableofcontents

\section{\label{sec:level1}INTRODUCTION}
The impossibility of perfect quantum cloning \cite{wootters}, marks a distinguishing departure of quantum information from its classical counterpart. Being one of the pioneering \enquote{no-go theorems}, impossibility in cloning together with other prohibited protocols \cite{nogotheorems1,nogotheorems2,nogotheorems3,nogotheorems4} make quantum information processing all the more secure.
However, the \enquote{No Cloning Theorem} does not rule out the possibility of approximate cloning. An arbitrary quantum state $|\psi\rangle$ can be cloned with a fidelity as high as $\frac{5}{6}$ \cite{approximatecloning1,approximatecloning2, cloningbuzek, buzek}. In a seminal paper \cite{buzek} Buzek et al.  initiated the idea of approximate cloning. In the process, two types of quantum cloning machines (QCM) namely state independent and state dependent machines were introduced. The universal quantum cloning machine (UQCM) \cite{buzek}, as it was coined, copied all the pure states equally well. It was categorized as a state independent quantum cloning machine (SIQCM), as the fidelity is independent of the input state parameters. This machine copies every state with fidelity $ \frac{5}{6} $ which was later proved to be optimal \cite{buzekoptimal,buzekoptimal1}. In addition, to this there are also state dependent quantum cloning machine (SQCM) whose fidelity depends on the choice of our input state \cite{statedependentmachine}. \\

Furthermore, there also exists probabilistic quantum cloning machine, with which we can clone an unknown quantum state, secretly chosen from a certain set of linearly independent states with a non-zero probability of success p ($0 <$ p $< 1$ ) \cite{approximatecloning2,statedependentmachine}. In an optimal asymmetric Pauli quantum cloning machine \cite{optimalassymcloning} the two output states are not identical. In higher dimension, optimal asymmetric Heisenberg cloning machine is defined which becomes an optimal symmetric cloning machine when $p = q = \frac{1}{2}$, where $p$ and $q$ are the machine parameters.\\

Quantum entanglement \cite{quantumentanglement1,quantumentanglement2} which is also known as the inseparability of composite quantum states represents the highest degree of non classical correlation between two quantum systems in the physical world. It acts as a vital resource in quantum cryptography \cite{secretsharing} and information processing tasks like teleportation \cite{teleportation}, super dense coding \cite{superdensecoding}, entanglement swapping \cite{entanglementswapping2}, and remote entanglement distribution \cite{remoteentanglementdistribution}. Entanglement cannot be increased by local operations and classical communication. However, nonlocal unitary operations on the composite system can generate entanglement between separable states, a fact used in experimental generation of entangled states \cite{expgen}. However, there are separable states which preserve their separability under any nonlocal unitary action. Such states are known as absolutely separable states and their characterization is a significant problem in quantum computing especially in context to NMR quantum computing \cite{nmr}. Another important feature in entanglement theory is the existence of entangled states with positive partial transpose (PPTES) \cite{pptes1,pptes2}. These weakly entangled states are however useful in some quantum cryptographic protocols \cite{pptcryp}. \\

There are instances where quantum correlations in general can go beyond the idea of entanglement. In the last decade, quantum discord \cite{discord1,discord2} was introduced to quantify quantum correlations that goes beyond the notion of entanglement. It must be emphasized here that discord actually supplements the measure of entanglement that
can be defined on the system of interest and at the same time can also act as a resource \cite{discordasaresource}. \\

Like entanglement, quantum coherence is something which lies at the heart of quantum mechanics and often viewed as the measure of superposition of quantum states \cite{quantifycoherence}. Quantum coherence have been used for crucial processes like better cooling \cite{Bras, Lost} or work extraction in nano-scale thermodynamics. Coherence also played a part in quantum algorithms \cite{Anand, Li, Cas} and in quantifying wave-particle duality \cite{WaveParticleDuality1, WaveParticleDuality2, WaveParticleDuality3}. Biological processes \cite{band, wilde} further vindicated the status of superposition as a significant ingredient. Due to the significant use of these resources, it is imperative to generate more number of states with these resources from a few in the form of broadcasting. \\

In the present contribution, we study the problem of broadcasting of quantum resources in $2 \otimes 3$ and in general for $2 \otimes d$ dimensional systems. We provide non-broadcastable ranges for a general mixed state in $2 \otimes 3$ dimension. In particular, we find out the broadcasting ranges of maximally entangled mixed states (MEMS) and two parameter class of states (TPCS). Our protocol also generates absolutely separable states and PPTES. Further, we show the impossibility of optimal broadcasting of quantum correlations beyond entanglement (QCsbE) and quantum coherence in a general qubit-qudit mixed quantum state using local symmetric optimal Heisenberg quantum cloning machine (QCM). However, illustrations are provided in support of the non-optimal broadcasting of these resources.\\

This paper is planned as follows; in section II, we give a brief description of related concepts which will be useful in the subsequent sections of the article. In section III, we study the broadcasting of entanglement for general qubit-qutrit state. As an example, we have also considered the broadcasting of maximally entangled mixed states (MEMS) and two parameter class of states (TPCS). In section IV, we have given the proof of the impossibility of optimal broadcasting of QCsbE and coherence and examples to show the non-optimal broadcasting of these resources. Finally, we conclude in section V. A summary of the previous results in contrast to this work is given in Table (\hypersetup{linkcolor=green}\ref{table_11}\hypersetup{linkcolor=blue}).
\begin{widetext}

\begin{table}[h!]
\setlength{\tabcolsep}{0.3em}
  \begin{center}
    \caption{Summary of earlier results and the present work on broadcasting of entanglement, discord and coherence. NME, MEMS, TPCS, 2-qubit general, qubit-qutrit general and qubit-qudit general stand for nonmaximally entangled state, maximally entangled mixed state, two parameter class of states, general two qubit mixed state, general qubit-qutrit mixed state, and general qubit-qudit mixed state, respectively.}
    \vspace{0.2cm}
    \begin{tabular}{c|c|c|c|c}
      \textbf{System's Dimension} & \textbf{Resource state} & \textbf{Broadcasting of} & \textbf{Cloning operation} & \textbf{Author(s)}\\
      \hline
      2 $\otimes$ 2 & NME & Entanglement & Symmetric & Buzek \textit{et al.} and Hillery \cite{nogotheorems2, cloningbuzek}\\
      
      2 $\otimes$ 2 & NME & Entanglement & Symmetric & Bandyopadhyay \textit{et al.} \cite{bandyopadhyay} \\
      
      2 $\otimes$ 2 & NME & Entanglement & Asymmetric & Ghiu \cite{optimalassymcloning} \\
      
      2 $\otimes$ 2 & 2-qubit general & Entanglement and Discord & Symmetric & Chatterjee \textit{et al.} \cite{sourav_chat}\\
      
      2 $\otimes$ 2 & 2-qubit general & Coherence & Symmetric & Sharma \textit{et al.} \cite{udit_sharma} \\
      
      2 $\otimes$ 2 & 2-qubit general & Entanglement and Discord & Asymmetric & Jain \textit{et al.} \cite{aditya_jain} \\
      
      2 $\otimes$ 3 & qubit-qutrit general & Entanglement & Symmetric & This work \\
      
      2 $\otimes$ 3 & MEMS and TPCS & Entanglement & Symmetric & This work\\
      
      2 $\otimes$ d & qubit-qudit general & Discord & Symmetric & This work\\
      
      2 $\otimes$ d & qubit-qudit general & Coherence & Symmetric & This work\\
      
    \end{tabular}
  \end{center}
  \label{table_11}
\end{table}

\end{widetext}

\section{Useful Definitions And Concepts}
In this section, we give a brief introduction to various concepts which will be useful and related to the main theme of the article.
\subsection{General qubit-qudit mixed state}
In this paper, we have considered a general qubit-qudit mixed state as a resource state which is represented in
the canonical form as,
\begin{equation}\label{1}
\begin{split}
\rho = \frac{1}{2d}\Bigg(\mathbb{I}_2\otimes\mathbb{I}_{d}  + \sum_{i=1}^3 x_i\sigma_i\otimes\mathbb{I}_{d} + \sum_{j=1}^{d^{2}-1}y_j\mathbb{I}_{2}\otimes O_j \\
+ \sum_{i=1}^3\sum_{j=1}^{d^{2}-1}T_{ij}\sigma_i\otimes O_j \Bigg).
\end{split}
\end{equation}
Here $x_i = Tr[\rho(\sigma_i\otimes\mathbb{I}_{d})]$, $y_j = Tr[\rho(\mathbb{I}_{2}\otimes O_j)]$, $T_{ij} =Tr[\rho(\sigma_i\otimes O_j)]$, $\sigma_i$'s are $2 \times 2$ Pauli matrices and $O_j$'s are $ (d^2 - 1) $ linearly independent operators defining an operator basis for the $ d- $dimensional subsystem with an additional property $ Tr(O_iO_j) = \delta_{ij} $. Here, $\mathbb{I}_d$ is the identity matrix of order $d$. Tr is the trace operation on a given matrix and $ \delta_{ij} $ is the Kronecker delta symbol.\\

As an example, a general qubit-qutrit mixed entangled state $\rho_{12}$ is given by, 
\begin{equation}\label{2}
\begin{split}
\rho_{12} = \frac{1}{6}\Bigg(\mathbb{I}_6 + \sum_{i=1}^3 x_i\sigma_i\otimes\mathbb{I}_3 + \sum_{i=1}^8y_i\mathbb{I}_2\otimes G_i \\
+ \sum_{i=1}^3\sum_{j=1}^8 T_{ij}\sigma_i\otimes G_j \Bigg) = \big\{\vec{X},\vec{Y},T\big\},
\end{split}
\end{equation}
where $x_i = Tr[\rho (\sigma_{i}\otimes \mathbb{I}_3)]$, $y_i = Tr[\rho(\mathbb{I}_2\otimes G_i)]$, $T_{ij} =Tr[\rho(\sigma_{i} \otimes G_j)]$, $\sigma_{i}$'s are Pauli matrices and $G_j$'s are Gell-Mann matrices. $\vec{X}$, $\vec{Y}$ and $T$ are the Bloch vectors and the correlation matrix respectively.
\subsection{Entanglement detection criteria}
In order to test the separability of a given bipartite state, we generally use the Peres-Horodecki (\textbf{PH}) criteria \cite{ppt_criteria1, ppt_criteria2}. This criteria is necessary and sufficient condition for detection of entanglement for bipartite systems with dimension $2\otimes2$ and $2\otimes3$.
\subsubsection{\textbf{Peres-Horodecki (PH) criteria}}If at least one of the eigenvalues of a partially transposed density operator for a bipartite state $\rho$ defined as $\rho_{m \mu,\eta v}^{T}$ = $\rho_{m v,\eta \mu}$ turns out to be negative, then we can say that the state $\rho$ is entangled.
Equivalently, this criteria can be translated to the condition that determinant of at least one of the two matrices
\begin{equation}
\begin{split}
\renewcommand{\arraystretch}{1.5} % give some more room
& W_{3}=
\left(\begin{array}{@{}c|c@{}}
  W_{2} &
  \begin{matrix}
  \rho_{00,10} \\
  \rho_{00,11} 
  \end{matrix}
\\ \hline
  \begin{matrix}
  \rho_{10,00} & \rho_{11,00}
  \end{matrix}
  & \rho_{10,10}
\end{array}\right)
 \:\:\text{or}\\
\renewcommand{\arraystretch}{1.5} % give some more room
& W_{4}=
\left(\begin{array}{@{}c|c@{}}
  W_{3} &
  \begin{matrix}
  \rho_{01,10} \\
  \rho_{01,11} \\
  \rho_{11,10} 
  \end{matrix}
\\ \hline
  \begin{matrix}
  \rho_{10,01} & \rho_{11,01} & \rho_{10,11} 
  \end{matrix}
  & \rho_{11,11} 
\end{array}\right)\:\:\:\:\:\:\:\:\:\:\:\:\:\:\:\:\:\:\:\:\:\:\:\:\:\:
\label{eq:w3w4}
\end{split}
\end{equation}
is negative; with determinant of 
%\begin{equation}
$W_2=\begin{bmatrix}\rho_{00,00} & \rho_{01,00} \\
\rho_{00,01} & \rho_{01,01} \\
\end{bmatrix}$ being simultaneously non-negative.
\\

As \textbf{PH} criteria requires us to compute eigenvalues, it is not always computationally feasible to compute eigenvalues of the density matrix that have several variables as a argument. To overcome this problem, we have used another separability criteria in terms of Bloch parameters ($\vec{X},\vec{Y},T$) which is comparatively easier to compute. 

\subsubsection{\textbf{Separability criteria in terms of bloch parameters}}In order to check the separability, we have used separability criteria \cite{detectionofentanglement} in terms of Bloch sphere representation of two quantum mechanical systems. This criteria makes use of Ky Fan matrix norm. Let A be a matrix that belongs to $\mathbb{C}^{m \times n}$. The Ky Fan matrix norm is defined as the sum of singular values $\sigma_{i}$, 
\begin{equation}
    ||A||_{KF} = \sum_{i=1}^{min\{m,n\}} \sigma_{i} = Tr\sqrt{A^{\dagger} A}.
\end{equation}
\ \ \ This criteria states that if a bipartite state of $M \otimes N$ satisfies
\begin{equation}
\begin{split}
        \sqrt{\frac{2(M-1)}{M}}||X||_{2} + \sqrt{\frac{2(N-1)}{N}}||Y||_{2}\\ 
        +\sqrt{\frac{4(M-1)(N-1)}{(MN)}}||T||_{KF}\leq 1,
\label{separability_criteria_BS}
\end{split}
\end{equation}
then it is a separable state. Here $||.||_{2}$ is the Euclidean norm.\\
Therefore if a bipartite state of $2 \otimes 3$ dimension with Bloch representation \eqref{2} satisfies
\begin{equation}
    ||X||_{2} + \sqrt{\frac{4}{3}} ||Y||_{2} + \sqrt{\frac{4}{3}}||T||_{KF} \leq 1,
\label{separability_criteria_23}
\end{equation}
then it is a separable state. However, if the state violates this condition, we cannot conclude whether the state is separable or entangled. 

\subsection{Absolutely separable states}
In the resource theory of entanglement, LOCC (local operations and classical communication) are considered to be free operations, as entanglement does not increase under LOCC. Local unitaries are subsets of LOCC, and under their action, entanglement remains unchanged. However, nonlocal or global unitary operations are not free as they can turn a separable state into an entangled state. The CNOT operation is one of the fundamental global unitary operations which can change even a product state to an entangled state. \\

However, there are separable states which can preserve separability under any arbitrary global unitary operation. These states are termed as absolutely separable states \cite{absep1} . If we denote the set of separable states by $\mathbf{S}$ and absolutely separable states by $\mathbf{AS}$, then $\mathbf{AS}= \{ \sigma  : U \sigma U^{\dagger} \in \mathbf{S} ~~ \forall ~~  U \}  $. Here $U$ denotes an unitary operator. It has been proven in \cite{absep2} that a two qubit state is absolutely separable iff $\lambda_1 \le \lambda_3 + 2 \sqrt{\lambda_2 \lambda_4}$, where $\lambda_i$'s are the eigenvalues of the density matrix of the state arranged in descending order. Later on in \cite{absep3}, the condition was extended to states in $ 2 \otimes d$ dimensions. Another interesting feature of the absolutely separable states is that they form a convex and compact set within the set of separable states \cite{absep4}. 

\subsection{Bound entangled states}
Pure entangled states can be distilled from a large number of mixed entangled states for use in quantum information protocols . However, there are mixed entangled states from which no pure entangled state can be extracted. Subsequently, they came to be known as bound entangled states \cite{pptes1}.\\

It was noted that any entangled state which has a positive partial transpose is bound entangled (also known as undistillable) and literature is rich with examples of entangled states having positive partial transpose (PPTES) \cite{pptes2}. However the question whether a state which has a negative partial transpose(NPT) is bound entangled is still open.\\

Although a weaker form of entanglement, PPTES have found utility in information protocols like quantum key generation \cite{pptcryp}. Therefore, both from a mathematical and physical perspective, generation of PPTES is an intriguing problem in quantum information science. Most of the constructions of PPTES have been through mathematical rigour, the number of physical construction being rare.\\

The \textit{realignment criteria} is one of the simplest tests that can detect entanglement in PPT states. It states that all separable states $\rho \in M_{m} \otimes M_{n}$ satisfy $||R(\rho)||_{tr} \leq 1$, where R : $M_{m} \otimes M_{n} \rightarrow M_{m,n} \otimes M_{m,n}$ is the linear "realignment" map defined on elementary tensors by R($\lvert i \rangle \langle j \rvert \otimes \lvert k \rangle \langle l \rvert$) = $\lvert i \rangle \langle k \rvert \otimes \lvert j \rangle \langle l \rvert$. $||R(\rho)||_{tr} > 1$, is a signature of the entanglement of $\rho$. 
\vspace*{-5mm}
\subsection{Cloning} As stated previously, the no-cloning theorem states that given an arbitrary quantum state $|\psi\rangle$, there doesn't exist complete positive trace preserving map (CPTP) $C$ that can transform a single copy of $|\psi\rangle$ to two copies of $|\psi\rangle$ i.e. $ C : |\psi\rangle \not\rightarrow  |\psi\rangle\otimes|\psi\rangle$. \\

In our work, we are interested in symmetric $1\rightarrow 1 + 1$ cloning machines. We use the symmetric ($p = q = \frac{1}{2}$) version of optimal universal asymmetric Heisenberg cloning machine, where $p$ and $q$ are machine parameters. This machine creates the second clone with maximal fidelity for a given fidelity of first one. The general unitary transformation for cloning of qudit by this machine is given by : 
\begin{eqnarray}
\begin{aligned}
\label{heisenberg}
U \lvert j\rangle_a\lvert 00\rangle_{bc}  \rightarrow \sqrt{\frac{2}{d+1}}\Big( &\lvert j\rangle_{a}\lvert j\rangle_{b}\lvert j\rangle_{c}\\ +\frac{1}{2}\sum_{r=1}^{d-1}\lvert j\rangle_{a} \lvert \overline{j+r}\rangle_{b}\lvert \overline{j+r}\rangle_{c} 
&+\frac{1}{2}\sum_{r=1}^{d-1}\lvert \overline{j+r}\rangle_{a} \lvert j\rangle_{b}\lvert \overline{j+r}\rangle_{c}\Big),
\label{cloning}
\end{aligned}
\end{eqnarray}
Here, suffixes '$a$' and '$b$' represent clones, '$c$' represents the ancillary state and '$d$' denotes the dimension. 
\subsection{Broadcasting of quantum resources by cloning}
In this subsection, we give a brief exposure to the idea of broadcasting of resources with the help of cloning machines. It is known that entanglement, QCsbE and coherence can be used as a resource for a wide range of information processing tasks. Given that, there is always a necessity of creating more number of resource pairs with lesser resourcefulness from a single resource pair with higher degree of resourcefulness. The process of decomposing a resource pair to more number of resource pairs is called broadcasting of quantum resources. We apply different strategies to do broadcasting of resources. One such strategy is to apply local cloning operations on each party subsystem sharing the resource. In the subsequent subsections, we describe how the broadcasting happens in qubit-qudit systems.
\subsubsection{\textbf{Broadcasting of entanglement}}
\begin{figure}[h]
\begin{center}
\[
\begin{array}{cc}
\includegraphics[scale=0.5]{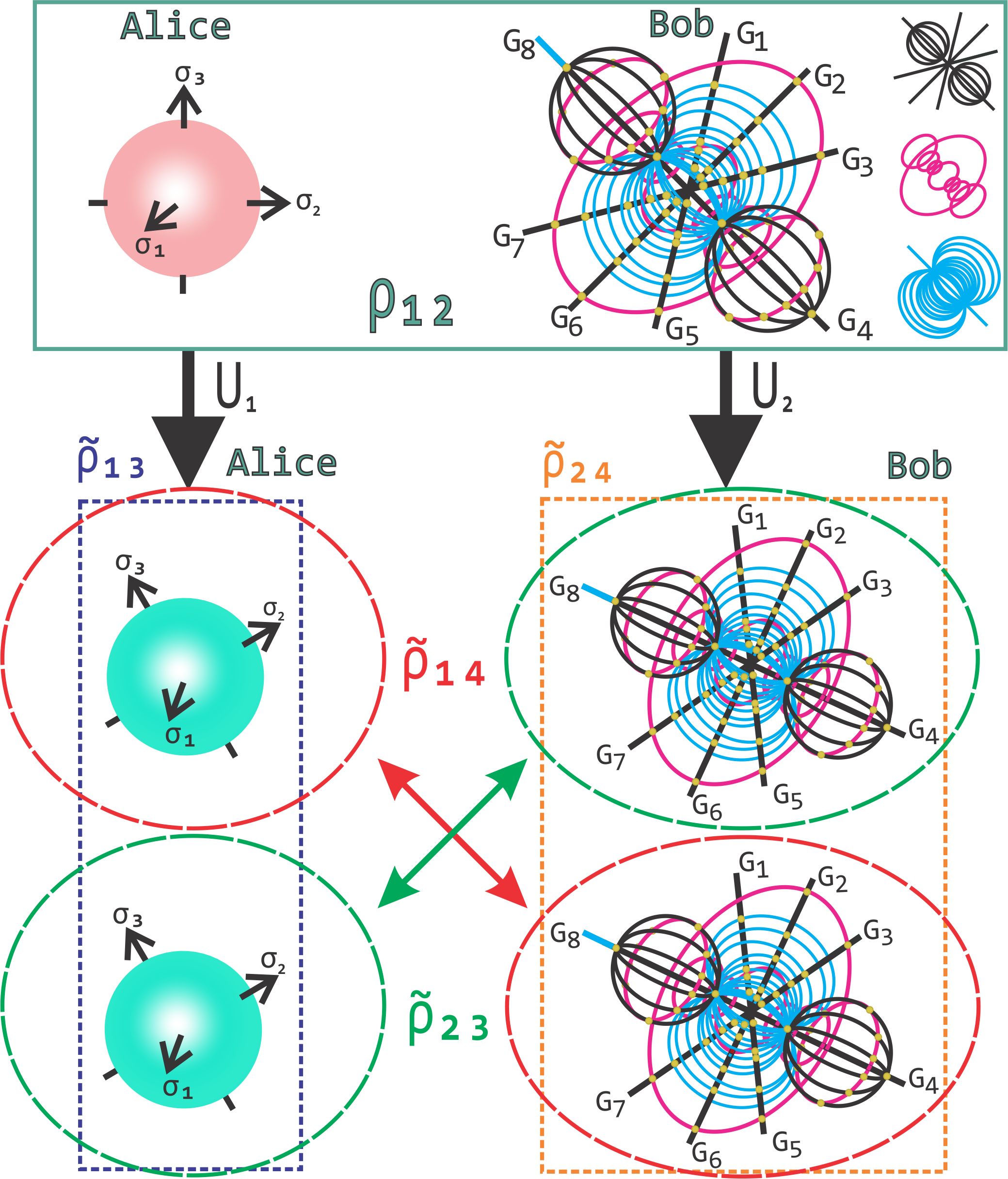}
\end{array}
\]
\end{center}
\caption{\noindent
%\scriptsize
A schematic depicting the application of local cloning unitaries $U_1$ and $U_2$ on a qubit-qutrit state shared between Alice \& Bob. The qutrit system ($d=3$) on Bob's side is illustrated with a 3-sphere or glome \cite{glome} structure in the four dimensional Euclidean space. Stereographic projection of the hypersphere's parallels (magenta), meridians (cyan) and hypermeridians (black) are illustrated separately, as subfigures on the right column, beside the glome qutrit for clarity. The projection being conformal, the circular curves intersect each other orthogonally at the yellow points. The rectangular box with a solid olive green boundary represents the input state $\rho_{12}$; while the dotted rectangular envelopes in blue and orange highlight the cloned local output pairs $\tilde{\rho}_{13}$ and $\tilde{\rho}_{24}$ respectively. Further, the dotted oval-shaped envelopes in red and green depict the cloned nonlocal output pairs $\tilde{\rho}_{14}$ and $\tilde{\rho}_{23}$ respectively. \label{fig:local_ent} 
}
\end{figure}

Let us consider that Alice and Bob share a general qubit-qudit mixed quantum state $\rho_{12}$ \eqref{1} as an input state. Also, qubit $3$ and qudit $4$ serve as the initial blank state in Alice's and Bob's individual subsystem respectively. We apply local cloning unitaries $U_{1}\otimes U_{2}$ \eqref{heisenberg} on qubits ($1,4$) and qudits ($2,4$). Tracing out ancilla qubit and ancilla qudit on Alice's and Bob's side respectively, we get the output state as $\tilde{\rho}_{1234}$. We trace out the ($2, 4$) and ($1, 3$) subsystems to obtain the local output states $\tilde{\rho}_{13}$ on Alice’s side and $\tilde{\rho}_{24}$ on Bob’s side respectively. Similarly, after tracing out appropriate qubits and qudits from the output state, we obtain the two plausible groups of nonlocal output states $\tilde{\rho}_{14}$ and $\tilde{\rho}_{23}$. The process is illustrated in figure (\ref{fig:local_ent}). \\

The expression for nonlocal outputs states across the subsystems of Alice and Bob are given by,
\begin{eqnarray}
\begin{aligned}
 \tilde{\rho}_{14} & = Tr_{23}[\tilde{\rho}_{1234}] \\
 & = Tr_{23}[U_1\otimes U_2(\rho_{12}\otimes B_{34}\otimes M_{56})U_{2}^{\dagger}\otimes U_{1}^{\dagger}],\\
 \tilde{\rho}_{23} & = Tr_{14}[\tilde{\rho}_{1234}] \\
 &= Tr_{14}[U_1\otimes U_2(\rho_{12}\otimes B_{34}\otimes M_{56})U_{2}^{\dagger}\otimes U_{1}^{\dagger}],\\ 
\end{aligned}
\end{eqnarray}
while the expression for local output states within Alice's and Bob's individual subsystem are given by,
\begin{eqnarray}
\begin{aligned}
\tilde{\rho}_{13} & = Tr_{24}[\tilde{\rho}_{1234}] \\
& =  Tr_{24}[U_1\otimes U_2(\rho_{12}\otimes B_{34}\otimes M_{56})U_{2}^{\dagger}\otimes U_{1}^{\dagger}],\\
\tilde{\rho}_{24} & = Tr_{13}[\tilde{\rho}_{1234}] \\
& = Tr_{13}[U_1\otimes U_2(\rho_{12}\otimes B_{34}\otimes M_{56})U_{2}^{\dagger}\otimes U_{1}^{\dagger}].\\
\end{aligned}
\end{eqnarray}

Here $B_{34} = \lvert 00\rangle \langle 00\rvert$ and $M_{56} = \lvert 00\rangle \langle 00\rvert$ represent the initial blank state and machine state respectively.
The requirement to broadcast entanglement between the desired pairs ($1,4$) and ($2,3$), we need to maximize the entanglement between nonlocal pairs ($1,4$) and ($2,3$) irrespective of the local pairs ($1,3$) and ($2,4$). However for optimal broadcasting, we should ideally have no entanglement between local pairs, thereby thereby increasing the amount of entanglement between nonlocal pairs.\\

\noindent\textbf{Non-optimal broadcasting of entanglement : }An entangled state $\rho_{12}$ is said to be broadcast after the application of local cloning operation 
($U_1 \otimes U_2$), if the nonlocal output states \{$\tilde{\rho}_{14}$, $\tilde{\rho}_{23}$\} are inseparable for some input state parameters.\\

\noindent\textbf{Optimal broadcasting of entanglement : }An entangled state $\rho_{12}$ is said to be broadcast optimally after the application of local cloning operation ($U_1 \otimes U_2$), if the nonlocal output states  \{$\tilde{\rho}_{14}$, $\tilde{\rho}_{23}$\} are inseparable and the local output states \{$\tilde{\rho}_{13}$, $\tilde{\rho}_{24}$\} are separable for some input state parameters.\\

\noindent\textbf{Sub-optimal broadcasting of entanglement : } An entangled state $\rho_{12}$ is said to be broadcast sub-optimally after the application of local cloning operation ($U_1 \otimes U_2$) for some input state parameters if the following conditions simultaneously hold : 
\begin{flushleft}
1. The nonlocal output states  \{$\tilde{\rho}_{14}$, $\tilde{\rho}_{23}$\} are inseparable.\\
2. Only one of the local output states \{$\tilde{\rho}_{13}$, $\tilde{\rho}_{24}$\} is separable. \\
%3. The local output state on Bob's (Alice's) side $\tilde{\rho}_{24}$ ($\tilde{\rho}_{13}$) can have few entangled states with separable states after we apply \textbf{PH} separability criteria.
\end{flushleft}

In this article, for sub-optimal broadcasting, we have considered the inseparability of nonlocal output states ($\tilde{\rho}_{14}$, $\tilde{\rho}_{23}$) and separability of local output states on Alice's side ($\tilde{\rho}_{13}$).
\subsubsection{\textbf{Broadcasting of quantum correlations beyond entanglement (QCsbE)}}
In the last decade, it was observed that entanglement is not sufficient to encapsulate all quantum correlations. It was also observed that there are correlations that go beyond the notion of entanglement and these QCsbE can be used as a resource for some operational tasks as they allow us to do these tasks more efficiently that would not be possible by any classical means. It is therefore equally important to broadcast QCsbE from a pair of state to a larger number of states. In a recent work, we have shown how to broadcast QCsbE in $2 \otimes 2$ systems \cite{sourav_chat,aditya_jain}. In this article, we have chosen geometric discord ($D_{G}$) to quantify the QCsbE.\\

\textbf{Geometric Discord ($D_{G}$) : }The geometric measure of quantum discord $D_{G}$ is a quantifier of general non-classical correlations in bipartite quantum states. It is the distance between the quantum state and the nearest classical state. For an arbitrary general qubit-qudit state $\rho_{12}$ (shared by parties numbered 1 and 2), it is defined as, 
$D_{G}(\rho_{12}) = \min_{\chi} || \rho_{12} - \chi ||^2,$ where $\chi$ is classical state. Such a classical state, in general, can be written as
$\chi = \sum_{i}^{d_{1}}p_{i}\pi_{i}^{1}\otimes\rho_{i}^{2} ,$
where $d_{1}$ is the dimension of subsystem 1 and $\pi_{i}^{1}$ are its projectors. $\rho_{i}^{2}$ are density matrices describing states of subsystem 2.\\

However, for a arbitrary qubit-qudit system, an analytical expression of $D_{G}$ has been obtained \cite{2xddiscord}, which is defined as follows : 
\begin{equation}\label{6}
\begin{aligned}
D_{G}(\rho_{12}) = \frac{1}{2d}(\lvert\lvert\vec{x}\rvert\rvert^2 + \lvert\lvert T\rvert\rvert^2 - \lambda_{\max}),
\end{aligned}
\end{equation} 
where $\vec{x}$ is the bloch parameter and $\lambda_{max}$ is the maximal eigenvalue of the matrix $\omega$ = ($\vec{x}\vec{x^t} + T T^t$). Here superscript 't' denotes the transpose and T is the correlation matrix of $\rho_{12}$.\\

Local broadcasting of QCsbE is very similar to the notion of local broadcasting of entanglement. Let $D_{G}$ be the total amount of QCsbE produced as a result of local cloning operations. $D_{G}^{l}$ and $D_{G}^{nl}$ represents the amount of QCsbE among local and across nonlocal parties, then $D_{G} = D_{G}^{l} + D_{G}^{nl}$. In order to maximize $D_{G}^{nl}$, $D_{G}^{l}$ should be ideally zero. \\

\noindent\textbf{Non-optimal broadcasting Of QCsbE : }A quantum correlated state $\rho_{12}$ is said to be broadcast after the application of local cloning operation 
($U_1 \otimes U_2$), if the amount of QCsbE of nonlocal output states \{$\tilde{\rho}_{14}$, $\tilde{\rho}_{23}$\} is non zero for some input state parameters.\\

\noindent\textbf{Optimal broadcasting Of QCsbE : }A quantum correlated state $\rho_{12}$ is said to be broadcast optimally after the application of local cloning operation ($U_1 \otimes U_2$), if the amount of QCsbE of nonlocal output states  \{$\tilde{\rho}_{14}$, $\tilde{\rho}_{23}$\} is non zero and the amount of QCsbE of local output states \{$\tilde{\rho}_{13}$, $\tilde{\rho}_{24}$\} is zero for some input state parameters.

%However in higher dimensions like d$\otimes$d ( where d$\geq$ 3), \textbf{PH} criteria is only a necessary condition but not a sufficient condition for separability. Therefore, there can be entangled states in higher dimensions like d$\otimes$d that remain positive under partial transposition. In our work, Bob's side dimension is d$\otimes$d.\\ 

%\noindent\textbf{Sub-Optimal Broadcasting Of QCsbE : }A quantum correlated state $\rho_{12}$ is said to be broadcast sub-optimally after the application of local cloning operation ($U_1 \otimes U_2$) for some input state parameters, if the amount of QCsbE of non local output states  \{$\tilde{\rho}_{14}$, $\tilde{\rho}_{23}$\} is non zero, the amount of QCsbE in Alice's local output states \{$\tilde{\rho}_{13}$, $\tilde{\rho}_{24}$\} is zero for some input state parameters but it does not give the entire range of input state parameters as the optimal one.  

\subsubsection{\textbf{Broadcasting of quantum coherence}}
Quantum Coherence has its application in variety of fields, ranging from quantum information processing to quantum sensing, metrology, thermodynamics \cite{Lost}, biology \cite{band} and it can act also as a resource in each of these domains. Therefore, it becomes important to investigate the possibility of creating more number of coherent states from an existing coherent pair. In a recent study, it has been shown that it is impossible to clone quantum coherence perfectly \cite{dhrumil}. In addition to this, just like entanglement, we have shown the possibility of broadcasting coherence using quantum cloning in $2 \otimes 2$ quantum system \cite{udit_sharma}. Due to the basis dependent property of quantum coherence, researchers have introduced the concept of genuine quantum coherence which are invariant of change of basis. In the process of cloning we have a blank state (suppose $\rho = \frac{\mathbb{I}}{2}$) which is genuinely incoherent state. So, if through the process of cloning, we try to increase coherence of the blank state, then the process is termed as broadcasting of quantum coherence.  Given a quantum state $\rho$, the amount of coherence present in the state $\rho$ in the basis ${|i\rangle}$ is given as follows,

\begin{equation}\label{7}
C(\rho) = \sum_{i \neq j} | \langle i | \rho | j \rangle | .
\end{equation}
We will calculate quantum coherence in the two-qubit computational basis ${|00\rangle, |01\rangle, |10\rangle, |11\rangle}$. This is $l_{1}$-norm and it does not depend upon diagonal elements and coherence will be zero in the eigenbasis of the density matrix.\\

To broadcast coherence between the desired pairs ($1,4$) and ($2,3$), one needs to maximize the amount of coherence between the nonlocal output pairs ($1,4$) and ($2,3$) irrespective of that between the local output pairs ($1,3$) and ($2,4$). In order to broadcast coherence optimally, the amount of coherence between local output pairs should be zero.\\

\noindent\textbf{Non-optimal broadcasting of coherence : }A coherent input state $\rho_{12}$ is said to be broadcast after the application of local cloning operation 
($U_1 \otimes U_2$), if the nonlocal output states \{$\tilde{\rho}_{14},\tilde{\rho}_{23}$\} are coherent i.e. C($\tilde{\rho}_{14}$)$\neq $0, C($\tilde{\rho}_{23}$)$\neq $0 for some input state parameters.\\

\noindent\textbf{Optimal broadcasting of coherence : }A coherent input state $\rho_{12}$ is said to be broadcast optimally after the application of local cloning operation ($U_1 \otimes U_2$), if the nonlocal output states \{$\tilde{\rho}_{14}$, $\tilde{\rho}_{23}$\} are coherent i.e. C($\tilde{\rho}_{14}$)$\neq $0 and C($\tilde{\rho}_{23}$)$\neq $0, while the local output states \{$\tilde{\rho}_{13}$, $\tilde{\rho}_{24}$\} are incoherent i.e. C($\tilde{\rho}_{13}$) = 0 and C($\tilde{\rho}_{24}$) = 0, for some input state parameters.

%\noindent\textbf{Sub-Optimal Broadcasting Of Coherence : }A coherent input state $\rho_{12}$ is said to be broadcast optimally after the application of local cloning operation ($U_1 \otimes U_2$), if the non local output states \{$\tilde{\rho}_{14}$, $\tilde{\rho}_{23}$\} are coherent i.e. C($\tilde{\rho}_{14}$)$\neq $0, C($\tilde{\rho}_{23}$)$\neq $0 ,while the local output states \{$\tilde{\rho}_{13}$, $\tilde{\rho}_{24}$\} are incoherent i.e. C($\tilde{\rho}_{13}$)=0, C($\tilde{\rho{24}}$)=0 for some input state parameters but it does not give the entire range of input state parameters as the optimal one.\\

\section{Broadcasting Of Entanglement In $2\otimes3$ Dimension}
In this section, we will demonstrate the broadcasting of entanglement for $2 \otimes 3$ system. Our input resource state is a general qubit-qutrit mixed state $\rho_{12}$ (as in Eq. \ref{2}). This state is shared between two parties, Alice and Bob. Both of them locally apply optimal universal symmetric Heisenberg cloning machine as given in Eq. \ref{heisenberg}.\\

After cloning, we trace out ancilla qubit and ancilla qutrit on Alice and Bob's side respectively. The state of this composite system is given by $\tilde{\rho}_{1234}$. We trace out 2,3 and 1,4 to get the nonlocal output states $\tilde{\rho}_{14}$ and $\tilde{\rho}_{23}$. Since we are using a symmetric cloner, both the nonlocal output states turns out to be same. \\

The expression of the reduced density operator for the nonlocal output states become : 
\begin{eqnarray}
\begin{aligned}
\tilde{\rho}_{14} = \tilde{\rho}_{23} = \bigg\{\frac{2}{3}\vec{X}, \frac{5}{8}\vec{Y}, \frac{5}{12} T \bigg\}.
\end{aligned}
\end{eqnarray}
\ \ \ Here $\vec{X}$ = $\{x_{i}\}_{i \in \{1,2,3\} }$, $\vec{Y}$ =  $\{y_{j}\}_{j \in \{1,..,8\} }$ and $T$ is the correlation matrix of the original input state.\\

\begin{figure}[h]
\begin{center}
\[
\begin{array}{cc}
\includegraphics[height=6cm,width=8cm]{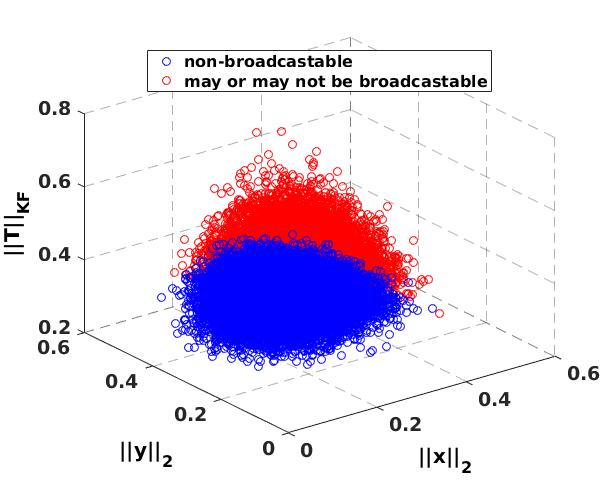}
\end{array}
\]
\end{center}
\caption{Blue colored circles denote the states that are not useful for broadcasting in $2 \otimes 3$ dimensions out of $ 5 \times 10^{4}$ states generated uniformly and randomly using Haar measure. Red colored circles mark the states that may or may not be useful for broadcasting in $2 \otimes 3$ dimensions.
%\scriptsize
\label{fig:state_generation} 
}
\end{figure}

Now, we need to apply entanglement detection criteria to check for the inseparability of nonlocal output states for non-optimal broadcasting. Since it is not computationally feasible to calculate the eigenvalues of the partial transpose of matrix $\tilde{\rho}_{14}$, we could not apply Peres-Horodecki criteria for this general case. As a result, we have used the separability criteria (Eq. \ref{separability_criteria_23}) in terms of Bloch parameters to check for the separability of nonlocal output states. By applying this criteria, we can only tell about the states which are not broadcastable. The non-broadcastable ranges are : 

\begin{eqnarray}
\begin{split}
\centering
\frac{2}{3}\sum_{j=1}^{8} \sqrt{\sum_{i=1}^{3} t_{ij}^{2} } &\leq \frac{12 - 8*A - 15*B}{10 \sqrt{3}},
\end{split}
\end{eqnarray}

where A = $\sqrt{\sum_{k=1}^{3} x_{k}^2}$, B = $\sqrt{\sum_{k=1}^{8} y_{k}^2}$ and $t_{ij}$ is the element of the $i^{th}$ row and the $j^{th}$ column of the correlation matrix.\\
\newline
To demonstrate the non-broadcastable states, we uniformly generated $5 \times 10^4$ states randomly using Haar measure. This is displayed in figure (\ref{fig:state_generation}). The states are represented by circles. The non-broadcastable states are shown in blue colored circles while the red colored circles shows the states that may or may not be useful for broadcasting.\\ 

To demonstrate the broadcasting of entanglement, we next consider two different classes of mixed entangled states, namely : (A) maximally entangled mixed states (MEMS) and (B) two parameter class of states (TPCS).

\subsection{Example : maximally entangled mixed states (MEMS)}
In this subsection, we consider maximally entangled mixed states (MEMS) as our first example to demonstrate broadcasting of entanglement. MEMS are states with maximum amount of entanglement for a given degree of mixedness. Its density matrix depends on the choice of  measures used to quantify entanglement and mixedness. Here, we use linear entropy as a measure for mixedness and square of concurrence as a measure of entanglement. For this choice,  the MEMS density matrices are divided into two sub-classes ($\rho^{MEMSI}$ and $\rho^{MEMSII}$) which are defined as follows \cite{MEMS} : 
\begin{equation}
\begin{split}
  \rho^{MEMS}=\begin{cases}
  \bigg( r \lvert\phi^{+}\rangle\langle\phi^{+}\lvert,\\ + \frac{1}{5} (1 + \frac{r}{2})(E_{2}+E_{5}),\\ + \frac{1}{5} (1 - 2r)(E_{1}+E_{3}+E_{6}) \bigg )  & r \in [0,\frac{1}{2}]\\
    r \lvert\phi^{+}\rangle\langle\phi^{+}\lvert + \frac{1}{2}(1 - r)(E_{2}+E_{5}) & r \in [\frac{1}{2},1].
  \end{cases}
 \end{split}
\end{equation}

Here, $E_{1}$=$\lvert 00\rangle\langle 00\lvert$, $E_{2}$=$\lvert 01\rangle\langle 01\lvert$, $E_{3}$=$\lvert 02\rangle\langle 02\lvert$, $E_{4}$=$\lvert 10\rangle\langle 10\lvert$, $E_{5}$=$\lvert 11\rangle\langle 11\lvert$, $E_{6}$=$\lvert 12\rangle\langle 12\lvert$, and $\lvert\phi^{+}\rangle$ = $\frac{1}{\sqrt{2}}(\lvert 00\rangle + \lvert12\rangle)$. We denote MEMS density matrix by $\rho^{MEMSI}$ and $\rho^{MEMSII}$ when r ranges from 0 to $\frac{1}{2}$ and $\frac{1}{2}$ to 1 respectively.
\subsubsection{MEMSI}We apply local optimal cloning transformations (as given in Eq. \ref{heisenberg}) for the subclass $\rho^{MEMSI}$. The expression for the reduced density operator of its nonlocal output states then become : 

\makeatletter 
\def\@eqnnum{{\normalsize \normalcolor (\theequation)}} 
\makeatother{\small\begin{eqnarray}\begin{split}
\tilde{\rho}_{14}^{MEMSI} &= \big\{\vec{X_{14}}^{MEMSI}, \vec{Y_{14}}^{MEMSI}, T_{14}^{MEMSI}\big\},\\    
\tilde{\rho}_{23}^{MEMSI} &= \big\{\vec{X_{23}}^{MEMSI}, \vec{Y_{23}}^{MEMSI}, T_{23}^{MEMSI}\big\},
\end{split} \end{eqnarray}} 

where $0\leq r \leq \frac{1}{2}$, $\vec{X_{14}}^{MEMSI}=\vec{X_{23}}^{MEMSI}=\{0,0,\frac{-2(-1+2r)}{15}\}$, $\vec{Y_{14}}^{MEMSI}=\vec{Y_{23}}^{MEMSI}=\{0,0,\frac{-2-r}{16},0,0,0,0,\frac{-2+9r}{16\sqrt{3}}\}$ and the non zero entries in the correlation matrix ($T_{14}^{MEMSI} = T_{23}^{MEMSI}$) are $t_{1,4} = \frac{5r}{12}, t_{2,5} = \frac{-5r}{12}, t_{3,3} = \frac{2+r}{24}$, and $t_{3,8} =\frac{2 + 11r}{24\sqrt{3}}$. Here, $t_{i,j}$ denotes the element in the $i^{th}$ row and the $j^{th}$ column of the correlation matrix. \\

%$$\begin{small}\vec{T_{nl}}^{MEMSI}\end{small} =  
%\left[ {\begin{array}{*{20}c}
%   0 & 0 & 0 & \frac{5r}{12} & 0 & 0 & 0 & 0 \\
%   0 & 0 & 0 & 0 & \frac{-5r}{12} & 0 & 0 & 0\\
%   0 & 0 & \frac{2+r}{24} & 0 & 0 & 0 & 0 & \frac{2 + 11r}{24\sqrt{3}}\\
% \end{array} } \right]
% $$
 
We now apply \textbf{PH} criteria to find out the condition for non-optimal broadcasting under which the nonlocal output states will be inseparable. We observe that the nonlocal output states are inseparable when the value of r is greater than $0.44$.\\

For optimal broadcasting, we need to check the separability of local output states along with the inseparability of nonlocal output states. The local output states for the input state $\rho^{MEMSI}$ are given by,
\makeatletter 
\def\@eqnnum{{\normalsize \normalcolor (\theequation)}} 
\makeatother{\small\begin{eqnarray}\begin{split}
\tilde{\rho}_{13}^{MEMSI} &= \big\{\vec{X_{13}}^{MEMSI}, \vec{Y_{13}}^{MEMSI}, T_{13}^{MEMSI}\big\},\\
\tilde{\rho}_{24}^{MEMSI} &= \big\{\vec{X_{24}}^{MEMSI}, \vec{Y_{24}}^{MEMSI}, T_{24}^{MEMSI}\big\},    
\end{split} \end{eqnarray}} 

where $\vec{X_{13}}^{MEMSI}$ = $\vec{Y_{13}}^{MEMSI}$ = $\{0,0,\frac{2-4r}{15}\}$, $T_{13}^{MEMSI}$ = $diag(\frac{1}{3},\frac{1}{3},\frac{1}{3})$, $\vec{X_{24}}^{MEMSI} = \vec{Y_{24}}^{MEMSI} = \{0,0,\frac{-2-r}{16},0,0,0,0,\frac{-2+9r}{16\sqrt{3}}\}$ and the non zero entries in the correlation matrix of Bob's side ($T_{24}^{MEMSI}$) are $t_{1,1} = \frac{6+3r}{40}, t_{2,2} = \frac{6+3r}{40}, t_{3,3} = \frac{6+3r}{40}, t_{4,4} = \frac{3-r}{20}, t_{5,5} = \frac{3-r}{20}, t_{6,6} = \frac{8-r}{40}, t_{7,7} = \frac{8-r}{40}, t_{8,8} = \frac{22-9r}{120},t_{3,8} = \frac{-2-r}{40\sqrt{3}}$, and $t_{8,3} = \frac{-2-r}{40\sqrt{3}} $. \\

%$$    
%\left[ {\begin{array}{*{20}c}
%   \frac{6+3r}{40} & 0 & 0 & 0 & 0 & 0 & 0 & 0\\
%   0 & \frac{6+3r}{40} & 0 & 0 & 0 & 0 & 0 & 0\\
%   0 & 0 & \frac{6+3r}{40} & 0 & 0 & 0 & 0 & \frac{-2-r}{40\sqrt{3}}\\
%   0 & 0 & 0 & \frac{3-r}{20} & 0 & 0 & 0 & 0\\
%   0 & 0 & 0 & 0 & \frac{3-r}{20} & 0 & 0 & 0\\
%   0 & 0 & 0 & 0 & 0 & \frac{8-r}{40} & 0 & 0\\
%   0 & 0 & 0 & 0 & 0 & 0 & \frac{8-r}{40} & 0\\
%   0 & 0 & \frac{-2-r}{40\sqrt{3}} & 0 & 0 & 0 & 0 & \frac{22-9r}{120}\\
% \end{array} } \right]
%$$

Now, for optimal broadcasting, we can apply \textbf{PH} criteria to check the separability of local output states and inseparability of nonlocal output states. Since \textbf{PH} criteria only provides a necessary condition for separability in $3 \otimes 3$ (Bob's side), there can be states that remain positive under partial transposition even if they are entangled. In this case, we therefore can give the sub-optimal broadcasting range under which the nonlocal output states are inseparable and local output state on Alice's side is separable. We observe that the local output states on Alice's side ($2 \otimes 2$) can be always separable irrespective of the value of 'r'. Therefore, the \textbf{sub-optimal} broadcasting range is same to the one obtained for \textbf{non-optimal} broadcasting.
\subsubsection{MEMSII}
We repeat the same procedure for MEMSII where 'r' lies between $\frac{1}{2}$ and 1. Its nonlocal output states are then given by, 
\makeatletter 
\def\@eqnnum{{\normalsize \normalcolor (\theequation)}} 
\makeatother{\small\begin{eqnarray}\begin{split}
\tilde{\rho}_{14}^{MEMSII} &= \big\{\vec{X_{14}}^{MEMSII}, \vec{Y_{14}}^{MEMSII}, T_{14}^{MEMSII}\big\},\\    
\tilde{\rho}_{23}^{MEMSII} &= \big\{\vec{X_{23}}^{MEMSII}, \vec{Y_{23}}^{MEMSII}, T_{23}^{MEMSII}\big\},
\end{split} \end{eqnarray}} 

where $\vec{X_{14}}^{MEMSII} = \vec{X_{23}}^{MEMSII}$ = $\{0,0,0\}$, $\vec{Y_{14}}^{MEMSII} = \vec{Y_{23}}^{MEMSII}$ = $\{0,0,\frac{15(-2+3r)}{32},0,0,0,0,\frac{-5\sqrt{3}(-2+3r)}{32}\}$ and the non zero elements in the correlation matrix ($T_{14}^{MEMSII} = T_{23}^{MEMSII}$) for nonlocal output states are $t_{1,4} = \frac{5r}{8}, t_{2,5} = \frac{-5r}{8}, t_{3,3} = \frac{5r}{16}$, and $t_{3,8} = \frac{5\sqrt{3}r}{16}$. Here, $t_{i,j}$ denotes the element in the $i^{th}$ row and the $j^{th}$ column of the correlation matrix. \\

%$$T_{nl}^{MEMSII} =  
%\left[ {\begin{array}{*{20}c}
%   0 & 0 & 0 & \frac{5r}{8} & 0 & 0 & 0 & 0 \\
%   0 & 0 & 0 & 0 & \frac{-5r}{8} & 0 & 0 & 0\\
%   0 & 0 & \frac{5r}{16} & 0 & 0 & 0 & 0 & \frac{5\sqrt{3}r}{16}\\
% \end{array} } \right]
%$$
 
We again apply \textbf{PH} criteria to find the condition of non-optimal broadcasting under which the nonlocal output states will be inseparable. We find out that nonlocal output states are always inseparable irrespective of any value of r between $\frac{1}{2}$ and 1. For optimal broadcasting, we check the separability of local output states along with the inseparability of nonlocal output states. The local output states for $\rho^{MEMSII}$ are given by, 
\makeatletter 
\def\@eqnnum{{\normalsize \normalcolor (\theequation)}} 
\makeatother{\small\begin{eqnarray}\begin{split}
\tilde{\rho}_{13}^{MEMSII} &= \big\{\vec{X_{13}}^{MEMSII}, \vec{Y_{13}}^{MEMSII}, T_{13}^{MEMSII}\big\},\\
\tilde{\rho}_{24}^{MEMSII} &= \big\{\vec{X_{24}}^{MEMSII}, \vec{Y_{24}}^{MEMSII}, T_{24}^{MEMSII}\big\},    
\end{split} \end{eqnarray}} 

where $\vec{X_{13}}^{MEMSII} = \vec{Y_{13}}^{MEMSII} = \{0,0,0\}$, $T_{13}^{MEMSII}$ = $diag(\frac{1}{3},\frac{1}{3},\frac{1}{3})$, $\vec{X_{24}}^{MEMSII}$ = $\vec{Y_{24}}^{MEMSII}$ = $\{0,0,\frac{5(-2+3r)}{16},0,0,0,0,\frac{5(2-3r)}{16\sqrt{3}}\}$ and the non zero entries in the correlation matrix of Bob's side ($T_{24}^{MEMSII}$) are $t_{1,1} = \frac{2-r}{8}, t_{2,2} = \frac{2-r}{8}, t_{3,3} = \frac{2-r}{8}, t_{4,4} = \frac{r}{4}, t_{5,5} = \frac{r}{4}, t_{6,6} = \frac{2-r}{8}, t_{7,7} = \frac{2-r}{8}, t_{8,8} = \frac{(2+3r)}{24},t_{3,8} = \frac{-2+3r}{8\sqrt{3}}$, and $t_{8,3} = \frac{-2+3r}{8\sqrt{3}}$. \\

%$$    
%\left[ {\begin{array}{*{20}c}
%   \frac{2-r}{8} & 0 & 0 & 0 & 0 & 0 & 0 & 0\\
%   0 & \frac{2-r}{8} & 0 & 0 & 0 & 0 & 0 & 0\\
%   0 & 0 & \frac{2-r}{8} & 0 & 0 & 0 & 0 & \frac{-2+3r}{8\sqrt{3}}\\
%   0 & 0 & 0 & \frac{r}{4} & 0 & 0 & 0 & 0\\
%   0 & 0 & 0 & 0 & \frac{r}{4} & 0 & 0 & 0\\
%   0 & 0 & 0 & 0 & 0 & \frac{2-r}{8} & 0 & 0\\
%   0 & 0 & 0 & 0 & 0 & 0 & \frac{2-r}{8} & 0\\
%   0 & 0 & \frac{-2+3r}{8\sqrt{3}} & 0 & 0 & 0 & 0 & \frac{(2+3r)}{24}\\
% \end{array} } \right]
%$$

We can apply \textbf{PH} criteria to check the separability of local output states. As stated earlier, there can be states in higher dimension like $3 \otimes 3$ that remains positive under partial transposition even if they are entangled since \textbf{PH} criteria only provides a necessary condition for $3 \otimes 3$ (Bob's side) dimension. In this case, we therefore can give the sub-optimal broadcasting range. We find out that the local output states on Alice's side ($2 \otimes 2$) can be separable when r $< 0.95$. Hence, we conclude that \textbf{non-optimal} broadcasting is always possible while for \textbf{sub-optimal} broadcasting, r should be less than $0.95$.\\

\textit{Absolutely separable states : } In what follows below, we show that our protocol generates absolutely separable states in Alice's side ($2 \otimes 2$) for some input state parameters. As noted earlier, absolutely separable states preserve their separability under global unitary operation.\\

For MEMSI, eigenvalues of the Alice's local output state are,  
$\lambda_{1} = \frac{6 - 2r}{15}, \lambda_{2} = \frac{1}{3}, \lambda_{3} = \frac{4 + 2r}{15}$, and $\lambda_{4} = 0$, such that the condition for absolute separability holds when r is exactly equal to $\frac{1}{2}$. \\ %\textcolor{red}{Purity = $\frac{77-8r+8 r^2}{225}$}\\

For MEMSII, eigenvalues of the Alice's local output state are, $\lambda_{1} = \frac{1}{3}, \lambda_{2} = \frac{1}{3}, \lambda_{3} = \frac{1}{3}$, and $\lambda_{4} = 0$, such that the condition for absolute separability holds for every r in range from $\frac{1}{2}$ to $1$. Therefore, we can say that absolute separability occurs in maximally entangled mixed state when r ranges from $\frac{1}{2}$ to $1$. \\
%\textcolor{red}{Purity = $\frac{1}{3}$}\\

\textit{PPT entangled states (PPTES) : } As noted earlier in Sec. II. D, the realignment criteria is used to detect entanglement in PPT states. By using this criteria, no PPTES in Bob's side ($3 \otimes 3$) were found for any input state parameter in MEMSI. On the other hand, PPTES states in Bob's side ($3 \otimes 3$) were found with MEMSII input states, when r ranges from $\frac{14 + 4\sqrt{6}}{25}$ to $1$. A typical PPTES in this range is given by, 
$$
\begin{bmatrix}
\frac{r}{4} & 0 & 0 & 0 & 0 & 0 & 0 & 0 & 0 \\
0 & \frac{2-r}{16} & 0 & \frac{2-r}{16} & 0 & 0 & 0 & 0 & 0 \\
0 & 0 & \frac{r}{8} & 0 & 0 & 0 & \frac{r}{8} & 0 & 0 \\
0 & \frac{2-r}{16} & 0 & \frac{2-r}{16} & 0 & 0 & 0 & 0 & 0 \\
0 & 0 & 0 & 0 & \frac{1-r}{2} & 0 & 0 & 0 & 0 \\
0 & 0 & 0 & 0 & 0 & \frac{2-r}{16} & 0 & \frac{2-r}{16} & 0 \\
0 & 0 & \frac{r}{8} & 0 & 0 & 0 & \frac{r}{8} & 0 & 0 \\
0 & 0 & 0 & 0 & 0 & \frac{2-r}{16} & 0 & \frac{2-r}{16} & 0 \\
0 & 0 & 0 & 0 & 0 & 0 & 0 & 0 & \frac{r}{4} \\
\end{bmatrix}
$$
These states will posses bound (or undistillable) entanglement. In the next subsection, we demonstrate the broadcasting of entanglement using our second example, a two parameter class of states (TPCS).
\subsection{Example : two parameter class of states (TPCS)}
We consider the following class of states with two real parameters $\alpha$ and $\gamma$ in 2 $\otimes$ 3 quantum system \cite{TPCScite} : 
\begin{equation}
\begin{split}
\rho_{\alpha,\gamma} = \alpha ( \lvert 02\rangle\langle 02\lvert +  \lvert 12\rangle\langle 12\lvert) + \beta(\lvert\phi^{+}\rangle\langle\phi^{+}\lvert + \lvert\phi^{-}\rangle\langle\phi^{-}\lvert \\ + \lvert\psi^{+}\rangle\langle\psi^{+}\lvert) + \gamma \lvert\phi^{-}\rangle\langle\psi^{-}\lvert,
\end{split}
\end{equation}
where $\lvert\phi^{\pm}\rangle$ = $\frac{1}{\sqrt{2}}(\lvert 00\rangle \pm \lvert 11\rangle)$ and $\lvert\psi^{\pm}\rangle$ = $\frac{1}{\sqrt{2}}(\lvert 01\rangle \pm \lvert 10\rangle)$ are the four bell states and the parameter $\beta$ is dependent on $\alpha$ and $\gamma$ by unit trace condition,
$2\alpha + 3\beta + \gamma = 1.$
From the unit trace condition, $\alpha$ can vary from $0$ to $\frac{1}{2}$ and $\gamma$ can vary from $0$ to $1$.\\

This input state is shared by two parties, Alice and Bob. They both apply local cloning transformations as given by Eq. \ref{heisenberg}. By tracing out the ancillas and appropriate qubit and qutrit on Alice's and Bob's side respectively, we get the nonlocal output states which are then given by, 
\makeatletter 
\def\@eqnnum{{\normalsize \normalcolor (\theequation)}} 
\makeatother{\small\begin{eqnarray}\begin{split}
\tilde{\rho}_{14} &= \big\{\vec{X_{14}}, \vec{Y_{14}}, T_{14}\big\},\\
\tilde{\rho}_{23} &= \big\{\vec{X_{23}}, \vec{Y_{23}}, T_{23}\big\},    
\end{split} \end{eqnarray}} 

where $\vec{X_{14}} = \vec{X_{23}}$ = $\{0,0,0\}$, $\vec{Y_{14}} = \vec{Y_{23}}$ = $\{0,0,0,0,0,0,0,\frac{15-90\alpha}{16\sqrt{3}}\}$ and the non zero entries in the correlation matrix ($T_{14} = T_{23}$) of nonlocal output states are $t_{1,1} = \frac{5-10\alpha-20\gamma}{24}, t_{2,2} =\frac{5-10\alpha-20\gamma}{24} $, and $t_{3,3} = \frac{5-10\alpha-20\gamma}{24}$. Here, $t_{i,j}$ denotes the element in the $i^{th}$ row and the $j^{th}$ column of the correlation matrix. \\

%$$\vec{T_{nl}} =  
%\left[ {\begin{array}{*{20}c}
%   \frac{5-10\alpha-20\gamma}{24} & 0 & 0 & 0 & 0 & 0 & 0 & 0\\
%   0 & \frac{5-10\alpha-20\gamma}{24} & 0 & 0 & 0 & 0 & 0 & 0\\
%   0 & 0 & \frac{5-10\alpha-20\gamma}{24} & 0 & 0 & 0 & 0 & 0\\
% \end{array} } \right]
%$$

Now, we apply the \textbf{PH} criteria to check the inseparability of these nonlocal output states for non-optimal broadcasting of entanglement. The \textbf{non-optimal} broadcasting is possible when the following condition is satisfied :\\
\begin{equation}
\label{non_optimal_TPCS}
\frac{31 - 50 \alpha - 40 \gamma}{96} < 0.    
\end{equation}

For optimal broadcasting, we also need to check the separability of local output states along with the inseparability of nonlocal output states. The local output states are given by, 
\makeatletter 
\def\@eqnnum{{\normalsize \normalcolor (\theequation)}} 
\makeatother{\small\begin{eqnarray}\begin{split}
\tilde{\rho}_{13} &= \big\{\vec{X_{13}}, \vec{Y_{13}}, T_{13}\big\},\\
\tilde{\rho}_{24} &= \big\{\vec{X_{24}}, \vec{Y_{24}}, T_{24}\big\},    
\end{split} \end{eqnarray}} 
where $\vec{X_{13}}$ = $\vec{Y_{13}}$ = $\{0,0,0\}$, $T_{13}$ = $diag(\frac{1}{3},\frac{1}{3},\frac{1}{3})$, $\vec{X_{24}}$ = $\vec{Y_{24}}$ = $\{0,0,0,0,0,0,0,\frac{5-30\alpha}{8\sqrt{3}}\}$ and the correlation matrix on Bob's side ($T_{24}$) is $diag(\frac{1-2\alpha}{4},\frac{1-2\alpha}{4},\frac{1-2\alpha}{4},\frac{1+2\alpha}{8},\frac{1+2\alpha}{8},\frac{1+2\alpha}{8},\frac{1+2\alpha}{8},\frac{1+6\alpha}{12})$.\\

%$$ \left[ {\begin{array}{*{20}c}
%   \frac{1-2\alpha}{4} & 0 & 0 & 0 & 0 & 0 & 0 & 0\\
%   0 & \frac{1-2\alpha}{4} & 0 & 0 & 0 & 0 & 0 & 0\\
%   0 & 0 & \frac{1-2\alpha}{4} & 0 & 0 & 0 & 0 & 0\\
%   0 & 0 & 0 & \frac{1+2\alpha}{8} & 0 & 0 & 0 & 0\\
%   0 & 0 & 0 & 0 & \frac{1+2\alpha}{8} & 0 & 0 & 0\\
%   0 & 0 & 0 & 0 & 0 & \frac{1+2\alpha}{8} & 0 & 0\\
%   0 & 0 & 0 & 0 & 0 & 0 & \frac{1+2\alpha}{8} & 0\\
%   0 & 0 & 0 & 0 & 0 & 0 & 0 & \frac{1+6\alpha}{12}\\
% \end{array} } \right]
%$$

Similar to the case of MEMS, we can only give the sub-optimal range as \textbf{PH} criteria is only a necessary condition for $3 \otimes 3$ dimension. The \textbf{sub-optimal} broadcasting is only possible when the following inequality is satisfied along with Eq. \ref{non_optimal_TPCS}: 
\begin{equation}
\frac{3 + 2 \alpha - \sqrt{11 - 76 \alpha + 204 \alpha^{2}}}{16} \geq 0.
\end{equation}

\begin{figure}
\begin{center}
\[
\begin{array}{cc}
\includegraphics[height=9cm,width=9cm]{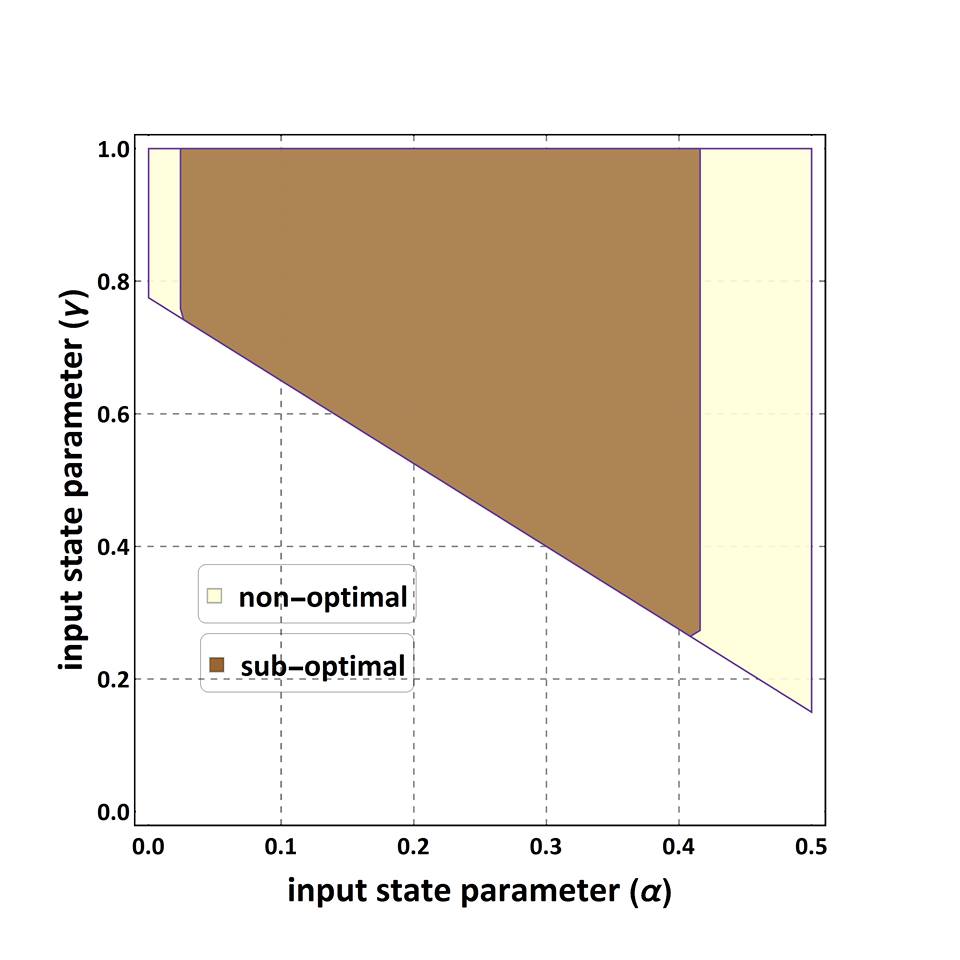}
\end{array}
\]
\end{center}
\caption{\noindent
%\scriptsize
Plot depicting the sub-optimal (in dark brown) as well as non-optimal (in light yellow) broadcastable region for the input TPCS in terms of two input state parameters : $\alpha$ and $\gamma$.}
\label{fig:tpcs_broadcasting} 
\end{figure}

In figure (\ref{fig:tpcs_broadcasting}), we depict the sub-optimal (in dark brown) and non-optimal (in light yellow) broadcastable regions when the input state is parameterized by $\alpha$ and $\gamma$. We observe that the broadcastable region for sub-optimal case is smaller as compared to non-optimal one due to extra separability constraint added on Alice's side in sub-optimal broadcasting. \\  

\textit{Absolutely separable states : } Alike maximally entangled mixed states, our protocol generates absolutely separable states on Alice's side ($2 \otimes 2$) with input TPCS too. The eigenvalues of the local output state on Alice's side are, $\lambda_{1} = \frac{1}{3}, \lambda_{2} = \frac{1}{3}, \lambda_{3} = \frac{1}{3}$, and $\lambda_{4} = 0$, so the absolute separability condition hold over the entire range of input state parameters. \\ %\textcolor{red}{Purity = $\frac{1}{3}$}\\

\textit{PPT entangled states (PPTES) : } Similar to MEMSII, our protocol generates PPTES on Bob's side ($3 \otimes 3$) also with input TPCS. PPTES are found at the output when state parameter $\alpha$ ranges from $0$ to $\frac{11 - 4\sqrt{6}}{50}$ and $\frac{11 + 4\sqrt{6}}{50}$ to $\frac{1}{2}$. A typical PPTES in this range is given by, 

$$
\begin{bmatrix}
\frac{1-2\alpha}{4} & 0 & 0 & 0 & 0 & 0 & 0 & 0 & 0 \\
0 & \frac{1-2\alpha}{8} & 0 & \frac{1-2\alpha}{8} & 0 & 0 & 0 & 0 & 0 \\
0 & 0 & \frac{1+2\alpha}{16} & 0 & 0 & 0 & \frac{1+2\alpha}{16} & 0 & 0 \\
0 & \frac{1-2\alpha}{8} & 0 & \frac{1-2\alpha}{8} & 0 & 0 & 0 & 0 & 0 \\
0 & 0 & 0 & 0 & \frac{1-2\alpha}{4} & 0 & 0 & 0 & 0 \\
0 & 0 & 0 & 0 & 0 & \frac{1+2\alpha}{16} & 0 & \frac{1+2\alpha}{16} & 0 \\
0 & 0 & \frac{1+2\alpha}{16} & 0 & 0 & 0 & \frac{1+2\alpha}{16} & 0 & 0 \\
0 & 0 & 0 & 0 & 0 & \frac{1+2\alpha}{16} & 0 & \frac{1+2\alpha}{16} & 0 \\
0 & 0 & 0 & 0 & 0 & 0 & 0 & 0 & \alpha \\
\end{bmatrix}
$$

\section{\label{sec:level1} Broadcasting of QCsbE and Coherence in $2\otimes d$ dimensions}
As discussed before, entanglement is not the only resource. There are correlations that go beyond entanglement : quantum discord. Other than quantum correlations that go beyond entanglement (QCsbE) (quantum discord), quantum coherence ($l_{1}$-norm) is also extensively used as a resource. Hence, a resource theory framework is created for describing them \cite{resource_theory_coherence}. In this section, we consider broadcasting of these resources in qubit-qudit system, where one of the parties say Alice is having a two-level system whereas other party in general is having a d-level system. The goal is again to create more number of resource states through broadcasting using optimal universal symmetric Heisenberg cloning machine. In this process, we find out that it is impossible to broadcast these resources optimally in a qubit-qudit system. However non-optimal broadcasting can still be done and we exemplify such cases of broadcasting in this section.

\subsection{\label{sec:level1}Optimal broadcasting of QCsbE and coherence}
In this subsection, we show that optimal broadcasting of QCsbE and quantum coherence is not possible in ($2\otimes d$)-dimensional system.\\

\noindent\textbf{Theorem 1 : } \textit{Given a general bipartite mixed quantum state in  $2\otimes d$ dimension $\rho_{12}$ (Eq. \ref{1}) and Heisenberg local cloning transformations (Eq. \ref{cloning}), it is impossible to broadcast the quantum discord ($D_{G}$ as defined in Eq. \ref{6}) within $\rho_{12}$ optimally into two lesser quantum correlated states: $\{\tilde{\rho}_{14}$, $\tilde{\rho}_{23}\}$.}  \\

\noindent\textbf{Proof:} Let us assume that two parties Alice and Bob share a general qubit-qudit quantum mixed state $\rho_{12}$. We then apply local Heisenberg optimal cloning transformations \eqref{heisenberg} to qubits '1' and '3' and qudit '2' and '4' on Alice's and Bob's side respectively. '5' and '6' are the machine state on Alice's and Bob's side respectively. By tracing out the machine states and Bob's side qudits, we get the local output state on Alice's part as $\tilde{\rho}_{13}$ = $\{\frac{2}{3}\vec{x}$ ,$\frac{2}{3}\vec{x}$ ,$T^{13}$\}, where $T^{13}$ = diag($\frac{1}{3},\frac{1}{3},\frac{1}{3}$) and $\vec{x} = \{x_{i}\}_{i \in \{1,2,3\}}$. We observe that the local output state on Alice's side does not depend on the dimension d of Bob's side (See Appendix A). The geometric discord $D_G$ calculated using Eq. \ref{6} of the local output state comes out to be constant i.e. $D_G$($\tilde{\rho}_{13}$) = $\frac{1}{18}$ which always remains non-zero. For optimal broadcasting, we need the $D_G$($\tilde{\rho}_{13}$) and $D_G$($\tilde{\rho}_{24}$) both to be zero. Hence optimal broadcasting of quantum discord is not possible in case of qubit-qudit system as $D_G$($\tilde{\rho}_{13}$) $\neq$ 0. \\

\noindent\textbf{Theorem 2 : }\textit{Given a general qubit-qudit mixed quantum state $\rho_{12}$ and Heisenberg optimal cloning transformations, it is impossible to broadcast the quantum coherence optimally within $\rho_{12}$ into two coherent states: $\{\tilde{\rho}_{14}$, $\tilde{\rho}_{23}\}$.}\\

\noindent\textbf{Proof:} We consider the input state shared between Alice and Bob as the most general qubit-qudit state $\rho_{12}$. We apply Heisenberg local cloning transformation $U_a \otimes U_b$ to clone the qubit $1 \rightarrow 3$ and qudit $2 \rightarrow 4$ on Alice and Bob's side respectively. By tracing out the machine states and Bob's side qudits, we get the local output state on Alice's part as $\tilde{\rho}_{13}$ = $\{\frac{2}{3}\vec{x}$ ,$\frac{2}{3}\vec{x}$ ,$T^{13}$\}, where $T^{13}$ = diag($\frac{1}{3},\frac{1}{3},\frac{1}{3}$) and $\vec{x} = \{x_{i}\}_{i \in \{1,2,3\}}$. The coherence given by $l_{1}$-norm (Eq. \ref{7}) of the local output state on Alice's side comes out to be $C(\tilde{\rho}_{13})=\frac{1}{3} + (\frac{4}{3})\sqrt{x_1^2+x_2^2} > 0$. For optimal broadcasting, we need the $C$($\tilde{\rho}_{13}$) and $C$($\tilde{\rho}_{24}$) both to be zero. Hence, it is evident that it is impossible to broadcast coherence optimally.

\subsection{Non-optimal broadcasting of QCsbE and coherence for MEMS and TPCS states}
In the previous subsection, we have seen that optimal broadcasting of quantum discord ($D_{G}$) and coherence ($l_{1}$-norm) is not possible for $2 \otimes d$ systems via optimal universal Heisenberg local cloning operations. However this never rules out the possibility of non-optimal broadcasting of these resources by using the same cloner. In this subsection we take the same qubit-qutrit examples: a) maximally entangled mixed states (MEMS) and b) two parameter class of states (TPCS) and show that non-optimal broadcasting is indeed possible to certain range of input state parameters. In particular, we find out the range based on the input state parameters for which such broadcasting will be possible.\\
\begin{figure}
\begin{center}
\[
\begin{array}{cc}
\includegraphics[height=4.5cm,width=9cm]{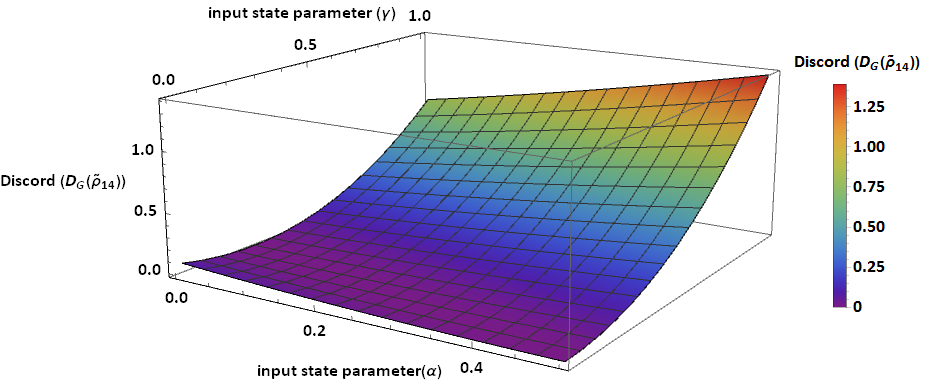}
\end{array}
\]
\end{center}
\caption{\noindent
%\scriptsize
The 3d-plot shows the variation of geometric discord ($D_{G}$) of nonlocal output state $\tilde{\rho}_{14}$ as a function of input state parameters $\alpha$ and $\gamma$ for the two parameter class of states.}
\label{fig:tpcs_discord} 
\end{figure}
\subsubsection{MEMS}
We apply local cloning transformation (Eq. \ref{heisenberg}) to MEMSI and MEMSII separately. We trace out the machine states and the respective qubits and qutrits to get the nonlocal output states ( $\tilde{\rho}_{14}$, $\tilde{\rho}_{23}$ ). Then, we calculate the geometric discord (Eq. \ref{6})of nonlocal output states. For non-optimal broadcasting, $D_{G}(\tilde{\rho}_{14})$ and $D_{G}(\tilde{\rho}_{23})$ is non zero for some input state parameter. In Table (\hypersetup{linkcolor=green}\ref{first_table}\hypersetup{linkcolor=blue}), we give the range for non-optimal broadcasting of geometric discord for the sub-classes : MEMSI and MEMSII.\\

For non-optimal broadcasting of coherence, we calculate the $l_{1}$-norm (Eq. \ref{7}) of nonlocal output states. Again for non-optimal broadcasting, $C(\tilde{\rho}_{14})$ and $C(\tilde{\rho}_{23})$ is non zero for some input state parameter. In Table (\hypersetup{linkcolor=green}\ref{second_table}\hypersetup{linkcolor=blue}), we give the range for non-optimal broadcasting of coherence for the sub-classes : MEMSI and MEMSII.

\begin{figure}
\begin{center}
\[
\begin{array}{cc}
\includegraphics[height=4.2cm,width=8cm]{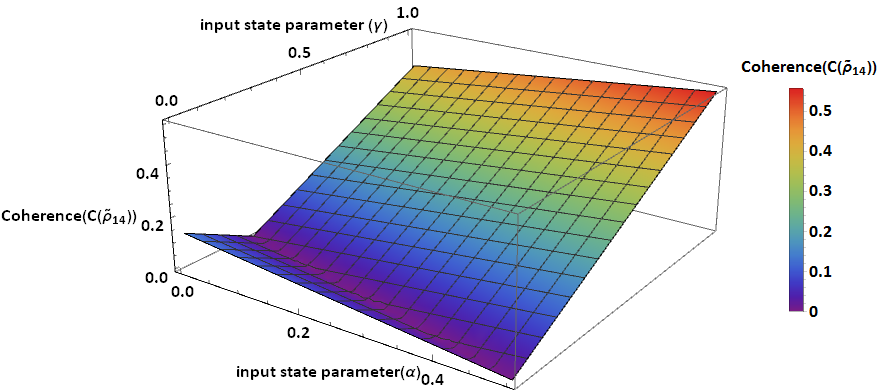}
\end{array}
\]
\end{center}
\caption{\noindent
%\scriptsize
The 3d-plot shows the variation of coherence ($l_{1}$-norm) of nonlocal output state $\tilde{\rho}_{14}$ as a function of input state parameters $\alpha$ and $\gamma$ for the two parameter class of states.}
\label{fig:tpcs_coherence}
\end{figure}

\subsubsection{TPCS}
We repeat the same procedure as above for two parameter class of states to find the range for non-optimal broadcasting of geometric discord (Eq. \ref{6}) in terms of input state parameters ($\alpha$ and $\gamma$). The expression for geometric discord comes out to be, 
 \begin{equation}
    D_{G}(\tilde{\rho}_{14}) = \frac{25 (-1 + 2\alpha + 4\gamma)^2}{288}. 
 \end{equation} 
We can clearly see that non-optimal broadcasting is possible for the entire range of $\alpha$ and $\gamma$ except for the points when $\alpha = \frac{1-4\gamma}{2}$. \\

\begin{table}
\begin{center}
    {\renewcommand{\arraystretch}{1.5}
    \begin{tabular}{ | c | c | c | c|}\hline
    States & $D_{G}(\tilde{\rho}_{14})$ & $D_{G}(\tilde{\rho}_{13})$ & Range\\ [2ex] 
    \hline\hline 
    MEMSI & $\frac{25 r^2}{192}$ & $\frac{1}{18}$ & $r>0$\\ [2ex]
    MEMSII & $\frac{25 r^2}{192}$ & $\frac{1}{18}$ & $r>0$\\ [2ex]
    \hline
    \end{tabular}}
    \newline\newline
    \caption{This table gives the range for non-optimal broadcasting of geometric discord ($D_{G}$) for MEMS class of states.}
\label{first_table}
\end{center}
\end{table}

\begin{table}
\begin{center}
    {\renewcommand{\arraystretch}{1.5}
    \begin{tabular}{ | c | c | c |c|}\hline
    States & $C(\tilde{\rho}_{14})$ & $C(\tilde{\rho}_{13})$ & Range\\ [2ex]
    \hline\hline
    MEMSI & $\frac{5 r}{12}$ & $\frac{1}{3}$ & $r>0$\\ [2ex] 
    MEMSII & $\frac{5 r}{12}$ & $\frac{1}{3}$ & $r>0$\\ [2ex]
    \hline
    \end{tabular}}
    \newline\newline
    \caption{This table gives the range for non-optimal broadcasting of coherence ($l_{1}$-norm) for MEMS class of states.}
\label{second_table}    
\end{center}
\end{table}

Though it is impossible to broadcast quantum coherence optimally but we can broadcast it non-optimally. We find the range for non-optimal broadcasting of coherence(Eq. \ref{7}) in terms of input state parameters ($\alpha$ and $\gamma$). The expression for coherence ($l_{1}$-norm) comes out to be, 
\begin{equation}
    C(\tilde{\rho}_{14}) = |\frac{5 - 10\alpha - 20\gamma}{36}|.
\end{equation}
\\
We can clearly observe that non-optimal broadcasting is possible for the entire range of $\alpha$ and $\gamma$ except for the points when $\alpha = \frac{1-4\gamma}{2}$. Broadcasting range for both discord and coherence with respect to input state parameters $\alpha$ and $\gamma$ are shown in the figure (\ref{fig:tpcs_discord}) and (\ref{fig:tpcs_coherence}) respectively. \\

\section{\label{sec:level1}CONCLUSION}
The present work deals with the broadcasting of quantum states beyond qubit-qubit systems. In particular, we investigate the problem of broadcasting of entanglement for a general  qubit-qutrit ($ 2 \otimes 3$) state and are able to identify the set of states for which the broadcasting will never be possible. We take examples like a) maximally entangled mixed states (MEMS) and b) two parameter class of states (TPCS) from $ 2 \otimes 3$ to show the range of both sub-optimal and non-optimal broadcasting. We show that it is impossible to optimally broadcast QCsbE and quantum coherence optimally for a general $ 2 \otimes d$ dimensional systems. Further to show that the non-optimal broadcasting of these resources is a still a possibility, we consider the same examples from $ 2 \otimes 3$ systems and thereafter find out the range of the input state parameters for which it will be possible.\\

Our protocol also results in states which are absolutely separable in two-qubit systems. Generation of entangled states having a positive partial transpose purely from physical consideration is another significant derivative of the work presented here. However, our work focuses on broadcasting of quantum resources in qubit-qutrit and qubit-qudit scenario. This work calls an attention for extension to arbitrary dimensions in bipartite and multipartite system. 

\section*{\label{sec:level1}ACKNOWLEDGEMENT}
N.G. would like to acknowledge support from the Research Initiation Grant of BITS-Pilani, Hyderabad vide letter no. BITS/GAU/RIG/2019/H0680 dated 22nd April, 2019.

\appendix
\section*{Appendix A}
Let us assume two parties Alice and Bob are sharing a general mixed state in $2 \otimes d$ dimension ($\rho_{12}$) as defined in Eq. \ref{1}. Both parties apply local optimal symmetric Heisenberg cloner on their respective sides. The blank state on Alice's and Bob's side are represented by suffix '3' and '4' respectively. The initial state of cloning machine state on Alice's side is denoted by '5' and that on Bob's side is denoted by '6'. The state of the composite system can be represented by $\rho_{123456}$. $U_a $ and  $U_b$ are the cloning operators on Alice's and Bob's side respectively. We then trace out '2' , '4', '6' subsystem from Bob's side, after the application of cloning machine to get, 
\begin{widetext}
\begin{equation}
    \begin{split}
       \rho_{135} &= Tr_{246}\Bigg[(U_a \otimes U_b) \rho_{123456} (U_a^{'} \otimes U_b^{'})\Bigg] \\
       &= Tr_{246}\Bigg[(U_a \otimes U_b)  \frac{1}{2d}\Big(\mathbb{I}_{2} \otimes \mathbb{I}_{d} +  \sum_{i=1}^3 x_i\sigma_{i}\otimes\mathbb{I}_{d} + \sum_{i=1}^{d^{2}-1}y_i\mathbb{I_2}\otimes O_i + \sum_{i=1}^3\sum_{j=1}^{d^{2}-1}t_{ij}\sigma_i\otimes O_j \Big) \otimes \rho_{35} \otimes \rho_{46}(U_a^{'} \otimes U_b^{'})\Bigg] \\
       &=  Tr_{246}\Bigg[(U_a \otimes U_b)  \frac{1}{2d}\Big(\mathbb{I}_{2} \otimes \mathbb{I}_{d} \otimes \rho_{35} \otimes \rho_{46} + \sum_{i=1}^3 x_i\sigma_{i}\otimes\mathbb{I}_{d} \otimes \rho_{35} \otimes \rho_{46} + \sum_{i=1}^{d^{2}-1}y_i\mathbb{I_2}\otimes O_i \otimes \rho_{35} \otimes \rho_{46}\\ 
       &\ \ \ \ \ \ \ \ \ \ \ + \sum_{i=1}^3\sum_{j=1}^{d^{2}-1}t_{ij}\sigma_i\otimes O_j \otimes \rho_{35} \otimes \rho_{46} \Big) (U_a^{'} \otimes U_b^{'})\Bigg] \\
       &=  Tr_{246}\Bigg[\frac{1}{2d}\Big(U_a( \mathbb{I}_{2} \otimes \rho_{35})U_a^{'} \otimes U_b(\mathbb{I}_{d} \otimes \rho_{46})U_b^{'} + \sum_{i=1}^3 x_i U_a( \sigma_{i} \otimes \rho_{35})U_a^{'} \otimes U_b(\mathbb{I}_{d} \otimes \rho_{46})U_b^{'} \\ 
       &\ \ \ \ \ \ \ \ \ \ \ + \sum_{i=1}^{d^{2}-1} y_i U_a( \mathbb{I}_{2} \otimes \rho_{35})U_a^{'} \otimes U_b(O_{i} \otimes \rho_{46})U_b^{'} + \sum_{i=1}^3\sum_{j=1}^{d^{2}-1} t_{ij} U_a( \sigma_i \otimes \rho_{35})U_a^{'} \otimes U_b(O_j \otimes \rho_{46})U_b^{'} \Big)\Bigg]\\
       &= \frac{1}{2d}\Bigg[Tr_{246}\Big[U_a( \mathbb{I}_{2} \otimes \rho_{35})U_a^{'} \otimes U_b(\mathbb{I}_{d} \otimes \rho_{46})U_b^{'}\Big]\Bigg] + \frac{1}{2d}\Bigg[Tr_{246}\Big[\sum_{i=1}^3 x_i U_a( \sigma_{i} \otimes \rho_{35})U_a^{'} \otimes U_b(\mathbb{I}_{d} \otimes \rho_{46})U_b^{'}\Big]\Bigg] \\
       &\ \ \ \ \ \ \ + \frac{1}{2d}\Bigg[Tr_{246}\Big[\sum_{i=1}^{d^{2}-1} y_i U_a( \mathbb{I}_{2} \otimes \rho_{35})U_a^{'} \otimes U_b(O_i \otimes \rho_{46})U_b^{'}\Big]\Bigg] + \frac{1}{2d}\Bigg[Tr_{246}\Big[\sum_{i=1}^3\sum_{j=1}^{d^{2}-1} t_{ij} U_a( \sigma_i \otimes \rho_{35})U_a^{'} \otimes U_b(O_j \otimes \rho_{46})U_b^{'} \Big]\Bigg]\\
    \end{split}
\end{equation}

%\begin{equation}
%    \begin{split}
%       \rho_{135} =  \frac{1}{2d}\Bigg[d * \Big[U_a( \mathbb{I}_{2} \otimes %\rho_{35})U_a^{'}\Big]\Bigg] + \frac{1}{2d}\Bigg[d * \Big[\sum_{i=1}^3 x_i U_a( \sigma_{i} \otimes \rho_{35})U_a^{'}\Big]\Bigg]
%    \end{split}
%\end{equation}
%& \rho_{135} =  \frac{1}{2}\Bigg[U_a( \mathbb{I}_{2} \otimes \rho_{35})U_a^{'} + \sum_{i=1}^3 x_i U_a( \sigma_{i} \otimes \rho_{35})U_a^{'}\Bigg]
\end{widetext}
The reduced density matrix on Alice's side is given by, $\rho_{135} =  \frac{1}{2}\Bigg[U_a( \mathbb{I}_{2} \otimes \rho_{35})U_a^{'} + \sum_{i=1}^3 x_i U_a( \sigma_{i} \otimes \rho_{35})U_a^{'}\Bigg]$ as unitary transformation doesn't affect the inner product of the system and  $\sigma_{i}$'s and $O_{j}$'s are traceless matrices and is independent of dimension of Bob's side. Therefore, on application of optimal universal Heisenberg local cloning transformations (Eq. \ref{heisenberg}) on a general bipartite mixed state in $2 \otimes d$ dimension (Eq. \ref{1}), the marginal state of Alice, remains independent of the dimension '$d$' of Bob's side.\\\\

\end{document}